\documentclass[aps,prd,reprint,superscriptaddress,nofootinbib]{revtex4-2}
\usepackage{amsmath,amssymb,bm,mathtools,mathrsfs}
\usepackage{hyperref}
\hypersetup{colorlinks=true,linkcolor=black,citecolor=black,urlcolor=black,
pdftitle={Black-Hole First Laws and Horizon Constraints: A Differential Rank Criterion},
pdfauthor={Gunn Kim}}

\begin{document}

\title{Black-Hole First Laws and Horizon Constraints: A Differential Rank Criterion}
\author{Gunn Kim}
\affiliation{Department of Physics and Astronomy, Sejong University, Seoul 05006, Republic of Korea}
\date{September 15, 2026}

\begin{abstract}
Thermodynamic rewritings of horizon equations depend on which relations are varied and which have already been imposed. We formulate a differential rank criterion that makes these choices explicit. The construction uses a stationary metric family together with an independently marked radial surface, physical charges determined by the metric parameters, and specified extensions of the horizon thermodynamic quantities. A first-law covector is pointwise representable by selected constraints precisely when it lies in the span of their differentials. We distinguish this algebraic condition from a smooth identity in a neighborhood. If regular constraints already define the complete physical family, pointwise representability follows from the first law itself; a neighborhood identity additionally depends on the chosen extensions. With explicit prescriptions, Kerr--Newman, rotating BTZ black holes in new massive gravity, and the new-type NMG black hole all exhibit exact alignment with their horizon constraints. By contrast, substituting an exact metric family into a field-equation component can annihilate that residual before any horizon condition is tested. We also give an explicit entropy--volume comparison and specify the role of fixed ensemble data in free-energy variations. The framework separates consequences of the established first law from additional claims about selected constraints and their extensions.
\end{abstract}

\maketitle
\raggedbottom

\section{Introduction}

Black-hole mechanics and Hawking radiation establish a thermodynamic relation between horizon geometry and conserved charges \cite{Bekenstein1973,Bardeen1973,Hawking1975}. The connection with gravitational dynamics takes several forms. Jacobson derived the Einstein equation from a local Clausius relation \cite{Jacobson1995}, whereas Padmanabhan and subsequent authors showed that horizon field equations can be reorganized into identities resembling the first law \cite{Padmanabhan2002,Paranjape2006,Kothawala2007,Kothawala2009}. These constructions motivate a careful accounting of the information used in a thermodynamic rewriting.

In a simple static geometry, multiplication of a horizon equation by a volume variation produces a natural pressure--volume term. Ahn \textit{et al.}\ instead use entropy variations to accommodate general parameter variations in several rotating and higher-derivative examples \cite{Ahn2026}. Such a rewriting uses more than a multiplier: it also uses the identification of temperature and entropy, definitions of conserved charges, and relations between the metric parameters and the horizon location. The differential content of these ingredients should be distinguished.

Two distinctions are particularly important. A horizon-location condition, such as $f(r_+)=0$, is different from a component of the gravitational field equations. Moreover, a field-equation residual evaluated on a family of exact solutions is different from the same residual before the field equations have been solved. Substituting an exact solution family can make a residual identically zero while leaving a horizon-location condition nontrivial on a space with an independently marked surface.

The covariant phase-space first law provides an established reference \cite{Wald1993,IyerWald1994}. For variations between stationary black holes at fixed couplings and appropriate boundary conditions, it relates the physical charges to the horizon temperature and gravitational entropy. We use this relation to test a specified set of reduced constraints. Our question is whether a chosen representative of the first-law covector can be written as a linear combination of their differentials. This is an elementary linear-algebra problem once the reduced space, constraints, and representative have been fixed.

The scope of the test requires equal care. If regular constraints already define the complete physical family, the first law implies the pointwise span condition. In that case a successful rank test is a consistency statement, not an independent derivation of thermodynamics. An exact identity in a surrounding neighborhood is stronger, but depends on how horizon quantities are extended away from the horizon condition. These observations delimit the physical interpretation of both success and failure.

We develop this distinction using a space of stationary metrics with a marked radial surface. Conserved charges remain functions of the metric parameters, and the surface becomes a physical horizon only after its location constraint is imposed. We obtain exact differential identities for Kerr--Newman, rotating BTZ in new massive gravity (NMG), and the new-type NMG black hole. We then compare horizon constraints with solved field-equation residuals, and discuss entropy, thermodynamic volume, and fixed-ensemble free energies. The contribution is a systematic formulation of the assumptions behind a reduced thermodynamic representation, rather than a new derivation of the black-hole first law or a universal criterion for all multiplier-based procedures.

\section{Physical first law and marked surfaces}
\subsection{Horizon-symmetry charge balance}

For a stationary axisymmetric black hole, the horizon Killing field has the form
\begin{equation}
\xi_H=\partial_t+\Omega_H\partial_\phi.
\end{equation}
In a gauge theory the relevant symmetry also includes a gauge transformation. The covariant phase-space construction equates the associated charge variations at infinity and at the horizon \cite{Wald1993,IyerWald1994}. With the usual charge and potential conventions, this gives
\begin{equation}
\delta M-\Omega_H\delta J-\Phi_H\delta Q
=T_H\delta S_{\rm grav}.
\label{eq:physicalfirstlaw}
\end{equation}
The symmetry parameters in this linearized identity are evaluated at the background solution. A field-dependent-generator formulation requires the corresponding adjusted charge variation; one must not add $J\delta\Omega_H$ or $Q\delta\Phi_H$ to Eq.~\eqref{eq:physicalfirstlaw} by differentiating an unadjusted generator.

Throughout, $\delta$ denotes a covariant phase-space variation and $d$ an exterior derivative on a finite-dimensional parameter space. The explicit examples have fixed couplings. Extended thermodynamics will be used separately to define the thermodynamic volume. The applications here are diffeomorphism covariant; gravitational Chern--Simons theories require a generalized entropy and charge construction \cite{Tachikawa2007}.

\subsection{Metric parameters and a surface label}

Let $\lambda=(\lambda^1,\ldots,\lambda^k)$ label stationary metrics $g(\lambda)$, with any matter fields understood. Fix the theory, asymptotic normalization, and boundary conditions for which the stated charges are defined and integrable. Introduce a radial surface $\Sigma_R$ and consider pairs
\begin{equation}
\mathcal P=\{(g(\lambda),\Sigma_R)\},
\qquad x=(R,\lambda^1,\ldots,\lambda^k).
\label{eq:markedspace}
\end{equation}
The coordinate $R$ labels the position of the marked surface in the specified radial coordinate. It is not an additional metric degree of freedom. Varying $R$ at fixed $\lambda$ moves the surface on which the prescribed quantities are evaluated.

A horizon constraint $C_H(R,\lambda)=0$, together with the choice of outer branch, selects the physical horizon family $\mathcal S_H\subset\mathcal P$. On that family $R=r_+(\lambda)$. In the examples below, $g(\lambda)$ remains an exact solution throughout $\mathcal P$: only the horizon-location condition is left unimposed. We therefore use ``away from the horizon constraint'' rather than treating these variations as off-shell variations of the bulk gravitational fields. A genuine off-shell field family could also be studied, but would be a different construction.

The conserved charges are $M(\lambda)$, $J(\lambda)$, and $Q(\lambda)$. A charge already determined by the metric and boundary data is not added as an independent coordinate. This is essential when comparing the information in metric residuals with that in charge variations.

\subsection{Extensions of horizon quantities}

We specify smooth functions $\bar S$, $\bar T$, $\bar\Omega$, and $\bar\Phi$ on $\mathcal P$ whose restrictions to $\mathcal S_H$ agree with the physical horizon quantities. A bar indicates this prescribed extension. Away from $C_H=0$, $\bar T$ need not be a Hawking temperature, nor need $\bar S$ be a physical entropy assigned to an arbitrary surface. For example, algebraically continuing a horizon-area formula does not necessarily give the area of a nonhorizon surface.

Define the extended first-law covector by
\begin{equation}
\bar\Theta=dM-\bar T\,d\bar S
-\bar\Omega\,dJ-\bar\Phi\,dQ.
\label{eq:extendedTheta}
\end{equation}
For the inclusion $\iota:\mathcal S_H\hookrightarrow\mathcal P$, the physical statement is
\begin{equation}
\iota^*\bar\Theta=0.
\label{eq:pullback}
\end{equation}
The charge construction fixes this relation on physical variations. It does not, by itself, select a unique extension to independently displaced marked surfaces. Each example below states its prescription explicitly.

\section{Rank criterion and its scope}
\subsection{Pointwise and smooth representations}

Let $C_A(x)=0$, $A=1,\ldots,m$, be selected constraints on $\mathcal P$. They may be horizon constraints or specified reduced field-equation residuals, but their origin must be stated. Write
\begin{equation}
(J_C)_{Ai}=\partial_i C_A,
\qquad \bar\Theta=\tau_i\,dx^i.
\end{equation}
At a point $p$, representation using these constraint differentials means
\begin{equation}
\bar\Theta|_p=\sum_A\mu^A_p\,dC_A|_p.
\label{eq:pointspan}
\end{equation}
This holds precisely when
\begin{equation}
\operatorname{rank}\begin{pmatrix}J_C\\\tau\end{pmatrix}
=\operatorname{rank}J_C.
\label{eq:rankcriterion}
\end{equation}
We call Eq.~\eqref{eq:pointspan} pointwise representability and define
\begin{equation}
\Delta_{\rm th}(p)=
\operatorname{rank}\begin{pmatrix}J_C\\\tau\end{pmatrix}
-\operatorname{rank}J_C\in\{0,1\}.
\label{eq:deficiency}
\end{equation}
Equivalently, $\Delta_{\rm th}=0$ if and only if
\begin{equation}
\bar\Theta(v)=0\quad\text{for all }v\in\ker dC_p.
\label{eq:kernel}
\end{equation}
Thus a vector satisfying $dC_A(v)=0$ but $\bar\Theta(v)\neq0$ identifies a direction missed by the selected constraints, relative to the chosen extension.

A smooth representation on a neighborhood $U$ requires finite smooth functions $\mu^A$ with
\begin{equation}
\bar\Theta=\sum_A\mu^A dC_A\quad\text{throughout }U.
\label{eq:smoothspan}
\end{equation}
If $J_C$ has constant rank on $U$ and Eq.~\eqref{eq:pointspan} holds at every point of $U$, such coefficients exist locally. At a rank-changing point, pointwise success does not imply smooth representability. For $C=x^3$ and $\bar\Theta=x\,dx$, the rank condition holds everywhere, including the origin, but the coefficient required for $x\neq0$ is $1/(3x)$ and diverges at the origin.

The index is conditional on $\mathcal P$, the selected residuals, and $\bar\Theta$. It is invariant under nonsingular coordinate changes. Under $\widetilde C=A(x)C$ with invertible $A$, the constraint-differential span is preserved on the common zero set, where
\begin{equation}
d\widetilde C=A\,dC.
\end{equation}
Away from that set, $d\widetilde C=A\,dC+(dA)C$, so an exact neighborhood identity is not automatically invariant under a field-dependent residual redefinition. Singular redefinitions can also destroy the pointwise span on the zero set; for example, $C$ and $C^2$ have the same zeros but generally different differentials there.

These statements concern representations generated by $dC_A$. They do not establish a no-go result for every procedure that multiplies a field equation by $dS$ or $dV$ and then introduces other identities. Any additional horizon or charge relation used in such a procedure must be included in the information being compared.

\subsection{When pointwise success is automatic}
\label{sec:automatic}

Suppose the selected regular constraints define the complete physical family locally:
\begin{equation}
\mathcal S_H=C^{-1}(0).
\end{equation}
Then
\begin{equation}
T_p\mathcal S_H=\ker dC_p,
\end{equation}
and the first law, Eq.~\eqref{eq:pullback}, implies Eq.~\eqref{eq:kernel}. Equivalently,
\begin{equation}
\bar\Theta|_p\in N_p^*\mathcal S_H
=\operatorname{span}\{dC_A|_p\},
\label{eq:conormal}
\end{equation}
where $N_p^*\mathcal S_H$ is the conormal space. For a single regular horizon condition, the span is one dimensional. Accordingly, pointwise success on the physical family in the three examples below is already implied by their established first laws.

There is a useful distinction between this result and Eq.~\eqref{eq:smoothspan}. For independent regular constraints, vanishing pullback implies a local expression
\begin{equation}
\bar\Theta=\sum_A\mu^A dC_A+\sum_A C_A\eta^A,
\label{eq:modconstraints}
\end{equation}
with smooth coefficients and smooth one-forms $\eta^A$. The second sum vanishes on the constraint surface as an ambient covector but need not vanish in its neighborhood. Appendix~\ref{app:geometry} gives the local argument. The physical first law therefore does not generally imply an exact neighborhood span identity.

If the selected constraints define a larger set than the physical family, Eq.~\eqref{eq:conormal} for the complete family does not guarantee Eq.~\eqref{eq:pointspan} for the selected subset. The rank test can then detect omitted differential directions. Such a test still requires a physically justified extension, particularly for directions transverse to the physical family. The present black-hole examples illustrate complete horizon constraints and do not supply a nontrivial failure for an independently specified incomplete sector.

\subsection{Dependence on the extension}
\label{sec:extension}

For a single horizon constraint, consider two entropy extensions that agree on $C_H=0$:
\begin{equation}
\bar S' = \bar S+hC_H,
\label{eq:entropyshift}
\end{equation}
where $h$ is smooth with the appropriate units. Holding the other extensions fixed gives
\begin{equation}
\bar\Theta'=\bar\Theta
-\bar T h\,dC_H-\bar T C_H\,dh.
\label{eq:extensionchange}
\end{equation}
Both representatives have the same pullback. If $\bar\Theta=\mu dC_H$, they remain aligned on the horizon constraint, but the $C_Hdh$ term can spoil exact alignment away from it. Thus an exact neighborhood identity is a statement about a specified continuation of thermodynamic data, not an invariant consequence of the physical first law alone.

The pointwise condition can be written as $\bar\Theta|_p\in\operatorname{Im}(dC_p^*)$ independently of coordinates. At regular points of the selected zero set, this image is the conormal space. At rank-changing points it remains defined even if the zero set is not a smooth submanifold. No symplectic or coisotropic condition is assumed in this finite-dimensional test.

\section{Three horizon-constraint identities}
\subsection{Kerr--Newman}

Use units $G=c=\hbar=k_B=1$. The metric parameters are $M,a,Q$, with physical angular momentum $J=aM$. On the marked-surface space $(R,M,a,Q)$, define
\begin{equation}
C_{\rm KN}=R^2-2MR+a^2+Q^2,
\qquad D=R^2+a^2.
\end{equation}
The outer horizon branch is selected by $C_{\rm KN}=0$ and, away from extremality, $R>M$. We prescribe the algebraic continuations
\begin{equation}
\bar S=\pi D,\quad
\bar T=\frac{R-M}{2\pi D},\quad
\bar\Omega=\frac{a}{D},\quad
\bar\Phi=\frac{QR}{D}.
\label{eq:KNextensions}
\end{equation}
Their restrictions are the standard horizon quantities. We do not identify all of them with physical observables on an arbitrary nonhorizon surface.

With $dJ=a\,dM+M\,da$, direct differentiation yields
\begin{align}
\bar\Theta^{\rm KN}
&=\frac{R}{D}\bigl[R\,dM-(R-M)dR-a\,da-Q\,dQ\bigr]\nonumber\\
&=-\frac{R}{2D}\,dC_{\rm KN}.
\label{eq:KNfactor}
\end{align}
This identity holds on the marked-surface space wherever the expressions are regular, without imposing $C_{\rm KN}=0$. The coefficient is smooth for $R>0$. The rank of $dC_{\rm KN}$ is one because its $dM$ coefficient is $-2R$. Hence $\Delta_{\rm th}=0$.

\subsection{Rotating BTZ in new massive gravity}

We use the NMG convention \cite{BHT2009PRL,BHT2009PRD,Nam2010}
\begin{align}
I_{\rm NMG}=\frac{1}{16\pi G}\int d^3x\sqrt{-g}\,
\biggl[&R_{\rm sc}+\frac{2}{\ell^2}\nonumber\\
&+\frac{1}{m^2}\biggl(R_{\mu\nu}R^{\mu\nu}
-\frac{3}{8}R_{\rm sc}^2\biggr)\biggr],
\label{eq:NMGaction}
\end{align}
where $R_{\rm sc}$ denotes the Ricci scalar, to distinguish it from the surface label $R$. The equations are
\begin{equation}
E_{\mu\nu}\equiv G_{\mu\nu}-\frac{1}{\ell^2}g_{\mu\nu}
+\frac{1}{2m^2}K_{\mu\nu}=0.
\label{eq:NMGeom}
\end{equation}
The bare scale $\ell$ and the effective AdS scale $L$ obey
\begin{equation}
\frac{1}{\ell^2}=\frac{1}{L^2}
\left(1-\frac{1}{4m^2L^2}\right).
\label{eq:scales}
\end{equation}
The rotating BTZ metric is
\begin{equation}
ds^2=-f(r)dt^2+\frac{dr^2}{f(r)}
+r^2\left(d\phi-\frac{J}{2r^2}dt\right)^2,
\end{equation}
with
\begin{equation}
f(r)=-M+\frac{r^2}{L^2}+\frac{J^2}{4r^2}.
\end{equation}
Here $M,J$ are metric parameters. In units $8G=1$, the physical charges are
\begin{equation}
\mathcal M=\alpha M,\qquad \mathcal J=\alpha J,
\qquad \alpha=1+\frac{1}{2m^2L^2}.
\end{equation}
On $(R,M,J)$, choose $C_{\rm BTZ}=f(R)$ and prescribe
\begin{equation}
\bar S=4\pi\alpha R,\qquad
\bar T=\frac{f'(R)}{4\pi},\qquad
\bar\Omega=\frac{J}{2R^2}.
\end{equation}
The couplings and hence $\alpha$ are fixed. It follows that
\begin{align}
\bar\Theta^{\rm BTZ}
&=d\mathcal M-\bar T\,d\bar S-\bar\Omega\,d\mathcal J\nonumber\\
&=\alpha\left[dM-f'(R)dR-\frac{J}{2R^2}dJ\right]\nonumber\\
&=-\alpha\,dC_{\rm BTZ}.
\label{eq:BTZfactor}
\end{align}
For $R>0$, $dC_{\rm BTZ}$ has rank one and the coefficient is regular. Thus $\Delta_{\rm th}=0$. At $\alpha=0$ the covector itself vanishes in this prescription, so the same index is obtained trivially and does not signify recovery of a nonzero thermodynamic direction.

\subsection{New-type NMG}

At the coupling
\begin{equation}
m^2=\frac{1}{2L^2},\qquad \ell^2=2L^2,
\end{equation}
NMG admits the static family
\begin{equation}
ds^2=-f(r)dt^2+\frac{dr^2}{f(r)}+r^2d\phi^2,
\qquad f(r)=\frac{r^2}{L^2}+\frac{br}{L}+c.
\label{eq:newmetric}
\end{equation}
The dimensionless metric parameters are $b,c$. In the charge convention of Ref.~\cite{Nam2010}, with $8G=1$,
\begin{equation}
\mathcal M(b,c)=\frac{b^2-4c}{2}.
\label{eq:newmass}
\end{equation}
We use the boundary conditions and charge prescription under which these parameter variations obey the stated first law; no independent hair work term is introduced. On the marked-surface space $(R,b,c)$, the selected constraint is
\begin{equation}
C_{\rm new}=F(R,b,c)
=\frac{R^2}{L^2}+\frac{bR}{L}+c.
\label{eq:newF}
\end{equation}
We prescribe
\begin{equation}
\bar S=4\pi(2R+bL),\qquad
\bar T=\frac{2R+bL}{4\pi L^2}.
\label{eq:newextensions}
\end{equation}
The entropy continuation follows by evaluating the static Wald-density expression at the marked radius, as detailed in Appendix~\ref{app:wald}; its physical entropy interpretation is required only at a Killing horizon.

On $F=0$, the roots satisfy $r_-=-bL-r_+$ and $2r_++bL=r_+-r_-$. Thus Eq.~\eqref{eq:newextensions} reduces to
\begin{equation}
S_{\rm grav}=4\pi(r_+-r_-),\qquad
T_H=\frac{r_+-r_-}{4\pi L^2}.
\end{equation}
Away from $F=0$, we use $2R+bL$ as an algebraic expression and do not interpret it as a separation between two horizons. The nonextremal outer branch has $R>0$ and $2R+bL>0$.

Since $d\mathcal M=b\,db-2\,dc$,
\begin{align}
\bar\Theta^{\rm new}
&=b\,db-2\,dc
-\frac{2R+bL}{L^2}(2\,dR+L\,db)\nonumber\\
&=-2\,dF.
\label{eq:newfactor}
\end{align}
The $dc$ coefficient of $dF$ is one, so the constraint has rank one and the representation is smooth. The new-type solution therefore also has $\Delta_{\rm th}=0$. No independent mass direction has been added.

\begin{table}[t]
\caption{Exact identities for the prescribed extensions on marked-surface spaces. Each constraint has rank one for the stated domain, and each has $\Delta_{\rm th}=0$. Pointwise alignment on the physical horizon family follows from the first law; the displayed neighborhood identities also use the specified extensions.}
\label{tab:identities}
\begin{ruledtabular}
\begin{tabular}{lll}
Family & Coordinates & $\bar\Theta=\mu\,dC_H$\\
\hline
Kerr--Newman & $(R,M,a,Q)$ & $\mu=-R/(2D)$\\
BTZ in NMG & $(R,M,J)$ & $\mu=-\alpha$\\
New-type NMG & $(R,b,c)$ & $\mu=-2$
\end{tabular}
\end{ruledtabular}
\end{table}

\section{Horizon constraints and solved field equations}
\label{sec:residuals}

The distinction between $C_H$ and $E^r{}_r$ can be checked directly. For the exact BTZ metric,
\begin{equation}
G^r{}_r=\frac{f'}{2r}+\frac{J^2}{4r^4}=\frac{1}{L^2},
\qquad K^r{}_r=-\frac{1}{2L^4}.
\end{equation}
Using Eq.~\eqref{eq:scales}, the full NMG residual is
\begin{equation}
E^r{}_r=\frac{1}{L^2}-\frac{1}{\ell^2}
-\frac{1}{4m^2L^4}\equiv0.
\label{eq:BTZzero}
\end{equation}
This holds at every radius for every member of the exact BTZ family. It does not require $f(R)=0$.

For the static ansatz in Eq.~\eqref{eq:newmetric}, the curvature component in the convention of Eq.~\eqref{eq:NMGeom} is \cite{Ahn2026}
\begin{equation}
K^r{}_r=-\frac{f'f'''}{4}+\frac{(f'')^2}{8}
+\frac{ff'''}{2r}-\frac{f'f''}{4r}.
\end{equation}
The quadratic solution has $f''=2/L^2$ and $f'''=0$, giving
\begin{equation}
K^r{}_r=\frac{1}{2L^4}-\frac{f'}{2rL^2}.
\end{equation}
At the new-type coupling,
\begin{equation}
E^r{}_r=\frac{f'}{2r}-\frac{1}{2L^2}
+L^2K^r{}_r\equiv0.
\label{eq:newzero}
\end{equation}
The cancellation in this example is already noted in Ref.~\cite{Ahn2026}. It is not a failure of the first law or a rank defect unique to the new-type family: the exact BTZ residual in Eq.~\eqref{eq:BTZzero} is also identically zero.

On these solution families, the bulk equations have already been satisfied. Their differentials along the family, including an independent displacement of the marked surface, therefore vanish. The horizon condition has a different task: it selects which marked surface is the horizon. Consequently, a nonzero $dC_H$ can coexist with $dE^r{}_r=0$ without any contradiction.

This also explains why the ordering of substitutions matters. An equation simplified using $C_H=0$ may be valid on the horizon while possessing a different continuation away from it. Reintroducing independent surface variations after that simplification does not automatically reconstruct the differential of the original residual. A rank comparison must retain the same underlying functions, or explicitly record the additional horizon relations used to replace them. The identities in Table~\ref{tab:identities} concern horizon constraints and make no claim that an identically solved bulk residual independently reproduces the first law.

\section{Entropy, volume, and free energy}
\subsection{Independent entropy and volume variations}

For a static one-parameter horizon, entropy, area, and volume are functions of the same radius; their nonzero differentials are locally proportional. This geometric fact helps explain the interchangeability of several multipliers in restricted examples. Rotation can remove this degeneracy. The independence of thermodynamic volume and area for rotating black holes is established in extended black-hole thermodynamics \cite{Kastor2009,Cvetic2011,Dolan2011}.

An explicit check is useful. In four-dimensional Kerr--AdS, with $G=1$, let $L$ be the AdS radius and $\Xi=1-a^2/L^2$. On the physical family, the entropy and thermodynamic volume are \cite{Cvetic2011,Dolan2011}
\begin{align}
S&=\frac{\pi(r_+^2+a^2)}{\Xi},\\
V_{\rm th}&=\frac{2\pi(r_+^2+a^2)
\bigl(2r_+^2L^2+a^2L^2-r_+^2a^2\bigr)}
{3L^2\Xi^2 r_+}.
\end{align}
The volume is conjugate to $P=3/(8\pi L^2)$ in $dM=T_HdS+\Omega_HdJ+V_{\rm th}dP$. We now hold $L$ fixed and vary $(r_+,a)$. Direct differentiation gives
\begin{equation}
dS\wedge dV_{\rm th}
=\frac{4\pi^2 a^3(r_+^2+a^2)(1+r_+^2/L^2)^2}
{3r_+^2\Xi^4}\,dr_+\wedge da.
\label{eq:SVwedge}
\end{equation}
For $r_+>0$, $0<|a|<L$, this is nonzero. It vanishes in the static limit. Equation~\eqref{eq:SVwedge} gives an explicit differential form of the established independence: entropy and thermodynamic volume cannot generally be substituted for one another as differential factors on a rotating state space.

\subsection{Fixed ensemble data}

Euclidean gravity provides a thermodynamic potential after the ensemble and its boundary terms are specified \cite{GibbonsHawking1977,IyerWald1995}. Let $T_B,\Omega_B,\Phi_B$ be external intensive data in the same normalization as the charges, and define on the chosen marked-surface space
\begin{equation}
\bar{\mathscr F}_B=M-T_B\bar S-\Omega_BJ-\Phi_BQ.
\end{equation}
At fixed external data,
\begin{equation}
d_x\bar{\mathscr F}_B
=dM-T_Bd\bar S-\Omega_BdJ-\Phi_BdQ,
\end{equation}
where $d_x$ varies only the coordinates of $\mathcal P$. At a physical equilibrium point $p$, with the external data matched to the horizon values,
\begin{equation}
\left.d_x\bar{\mathscr F}_B\right|_p=\bar\Theta|_p.
\label{eq:fixedF}
\end{equation}
This is an equality of covectors at that point for the stated extension. In particular, its pullback to allowed physical variations vanishes. It is not an assertion that all directions in the artificial extension describe gravitational equilibrium variations.

If the external data are varied as well, the total differential contains
\begin{equation}
d\bar{\mathscr F}_B=d_x\bar{\mathscr F}_B
-\bar S\,dT_B-J\,d\Omega_B-Q\,d\Phi_B.
\end{equation}
For a smooth Euclidean saddle and the appropriate ensemble, $I_E=\beta_B\mathscr F_B^{\rm eq}$ with $\beta_B=T_B^{-1}$. This on-shell relation does not identify the arbitrary continuation $\bar{\mathscr F}_B$ with a Euclidean action away from the saddle. The ensemble-dependent potential and the horizon-symmetry charge balance consequently serve related but distinct purposes.

\section{Discussion}

The rank criterion organizes a specified reduced representation of the first law. Its inputs are the metric and surface variables, the selected constraint functions, and the extensions of the horizon thermodynamic data. Each input matters. Adding a charge coordinate that does not correspond to an independent field or boundary variation creates a direction that metric residuals cannot detect. Substituting an exact solution into a field-equation residual can remove all of that residual's differential content. Changing an entropy extension can alter an exact neighborhood identity without changing any physical first-law variation.

For a regular constraint set that already defines the physical horizon family, the pointwise criterion is a reformulation of the first law in conormal language. This automaticity is useful for checking consistency, but limits what can be inferred from the three successful examples. Their explicit coefficients provide exact identities for the chosen continuations; they do not establish a new thermodynamic law. A stronger diagnostic application would select a physically motivated incomplete residual sector, define the transverse variations independently, and identify a missed direction by Eq.~\eqref{eq:kernel}. At singular points, the regularity of the coefficient functions must also be checked.

The relation to multiplier-based horizon thermodynamics is correspondingly precise. Such constructions may use field equations together with horizon relations and charge definitions. Our test asks whether the differential span of a stated subset contains a stated first-law representative. It cannot assign all the information in a successful rewriting to whichever equation happened to be multiplied last. Conversely, failure for one subset and one continuation cannot rule out a construction that uses additional relations.

The covariant phase-space first law remains the physical reference because it compares horizon and asymptotic charges for admissible variations. The bulk equations enter that variational identity before restriction to a solved finite-dimensional metric family. Nothing in the analysis replaces the charge construction with a purely local temperature identification, or makes a preferred entropy continuation a new physical observable. Entropy retains its horizon role, while free energies require ensemble data and thermodynamic volume provides an independent direction in rotating examples.

All three explicit black-hole families pass the same horizon-constraint test. The analysis separates a consequence of the first law on its physical domain from a stronger identity for prescribed extensions. This distinction makes the assumptions in a horizon-thermodynamic derivation assessable.

\section{Conclusions}

We formulated a differential rank test for representing an extended first-law covector by selected reduced constraint differentials. Pointwise representability is equivalent to an augmented-rank equality. A smooth neighborhood identity additionally requires regular coefficient functions and pointwise inclusion throughout the neighborhood.

When regular constraints define the complete physical horizon family, pointwise success follows directly from the established first law. The physical statement permits terms proportional to the constraints in a surrounding neighborhood; excluding such terms is a stronger condition that depends on the extensions of the horizon quantities. The distinction is explicit on spaces of stationary metrics with an independently marked radial surface.

For the prescriptions given here, Kerr--Newman, rotating BTZ in NMG, and new-type NMG satisfy
\begin{align}
\bar\Theta^{\rm KN}&=-\frac{R}{2D}dC_{\rm KN},\nonumber\\
\bar\Theta^{\rm BTZ}&=-\alpha dC_{\rm BTZ},\nonumber\\
\bar\Theta^{\rm new}&=-2dF.
\end{align}
The exact radial field-equation residuals of the solved BTZ and new-type metric families instead vanish identically. These statements concern different constraints and are mutually consistent. Entropy--volume independence and fixed-ensemble free-energy variations further clarify why a thermodynamic differential cannot be selected independently of the variation space. A meaningful sufficiency test must therefore state both the information supplied by the selected constraints and the continuation used to represent the first law.

\appendix
\section{Local differential structure}
\label{app:geometry}

The pointwise criterion follows from the duality identity
\begin{equation}
(\ker dC_p)^\circ=\operatorname{Im}(dC_p^*).
\end{equation}
In coordinates, the right-hand side is the row space of $J_C$. Appending $\tau$ leaves the rank unchanged exactly when $\tau$ lies in that space. On a constant-rank neighborhood, a nonvanishing minor selects a smooth local frame for the row space, yielding smooth coefficients whenever the inclusion holds throughout that neighborhood.

For Eq.~\eqref{eq:modconstraints}, choose independent regular constraints and use local coordinates $(u^1,\ldots,u^q,y^1,\ldots,y^{n-q})$ with $u^A=C_A$ and physical family $u=0$. Write
\begin{equation}
\bar\Theta=a_A(u,y)du^A+b_j(u,y)dy^j.
\end{equation}
Vanishing pullback means $b_j(0,y)=0$. Smoothness gives
\begin{align}
b_j(u,y)&=\sum_A u^A B_{Aj}(u,y),\\
B_{Aj}(u,y)&=\int_0^1\partial_{u^A}b_j(tu,y)\,dt.
\end{align}
Consequently,
\begin{equation}
\bar\Theta=\sum_A a_A\,dC_A
+\sum_A C_A\left(\sum_j B_{Aj}dy^j\right),
\end{equation}
which establishes Eq.~\eqref{eq:modconstraints}. The additional one-form terms disappear on the zero set, but need not disappear nearby. This argument assumes regularity; at singular points we retain the image of $dC_p^*$ rather than identifying it with the conormal of a possibly singular set.

\section{Static NMG entropy prescription}
\label{app:wald}

For the Lagrangian in Eq.~\eqref{eq:NMGaction}, contraction of its curvature derivative with the horizon binormal gives the Wald entropy factor \cite{Wald1993,IyerWald1994}
\begin{equation}
1+\frac{1}{m^2}
\left(R^t{}_t+R^r{}_r-\frac34 R_{\rm sc}\right)
\end{equation}
for the static circular ansatz. With $ds^2=-fdt^2+dr^2/f+r^2d\phi^2$,
\begin{equation}
R^t{}_t=R^r{}_r=-\frac12\left(f''+\frac{f'}r\right),
\qquad R_{\rm sc}=-f''-\frac{2f'}r.
\end{equation}
Thus, in units $8G=1$, the entropy at a Killing horizon is
\begin{equation}
S_{\rm grav}=4\pi r_+
\left[1+\frac{1}{m^2}
\left(-\frac{f''(r_+)}4+\frac{f'(r_+)}{2r_+}\right)\right].
\end{equation}
Our static prescription replaces $r_+$ by $R$ in this expression. For the new-type quadratic metric at $m^2=1/(2L^2)$, it gives
\begin{equation}
\bar S=4\pi R\left(2+\frac{bL}{R}\right)
=4\pi(2R+bL),
\end{equation}
as used in Eq.~\eqref{eq:newextensions}. This supplies a concrete continuation associated with the chosen Lagrangian expression. It does not assert a unique entropy for general nonhorizon surfaces; the dependence discussed in Sec.~\ref{sec:extension} remains relevant.

\end{document}